\documentclass[a4paper]{jpconf}
\usepackage{graphicx}
\usepackage{amsmath,amssymb}
\usepackage{latexsym}
\usepackage{epstopdf}
\usepackage{subcaption}
\usepackage{citesort}
\begin{document}
\begin{flushright}
WITS-MITP-021
\end{flushright}

\title{Absorption probabilities associated with spin-3/2 particles near $N$-dimensional Schwarzschild black holes}

\author{G. E. Harmsen$^{a, 1}$ C. H. Chen$^{b, 2}$, H. T. Cho$^{b, 3}$, and A. S. Cornell$^{a, 4}$}
\address{$^{a}$ National Institute for Theoretical Physics; School of Physics and Mandelstam Institute for Theoretical Physics, University of the Witwatersrand, Johannesburg, Wits 2050, South Africa}
\address{$^{b}$ Department of Physics, Tamkang University, Tamsui, Taipei, Taiwan}
\ead{$^{1}$gerhard.harmnen5@gmail.com, $^{2}$899210057@s99.tku.edu.tw, $^{3}$htcho@mail.tku.edu.tw, $^{4}$alan.cornell@wits.ac.za}
\begin{abstract}
In June 2015 the Large Hadron Collider was able to produce collisions with an energy of 13TeV, where collisions at these energy levels may allow for the formation of higher dimensional black holes. In order to detect these higher dimensional black holes we require an understanding of their emission spectra. One way of determining this is by looking at the absorption probabilities associated with the black hole. In this proceedings we will look at the absorption probability for spin-3/2 particles near $N$-dimensional Schwarzschild black holes. We will show how the Unruh method is used to determine these probabilities for low energy particles. We then use the Wentzel-Kramers-Brillouin approximation in order to determine these absorption probabilities for the entire possible energy range.
\end{abstract}

\section{Introduction}

\par If we consider a system containing only a single Schwarzschild black hole and a particle of energy $\omega$, such that $\omega$ is very small compared to the total energy of the black hole, then classically the Schwarzschild metric would sufficiently describe the system (as the particle can be considered to be point like). Using only the Schwarzschild metric we could then determine the probability that the particle is absorbed by the black hole. This probability, however, is vastly different to the one that is obtained using a field theoretic approach \cite{funkhouser2010particle}. In 1976 Unruh showed that even for low energy particles there is significant deviation in absorption probabilities for the classical case and the field theory case  \cite{unruh1976absorption}. In field theoretic models it is possible to have the particle reflected off the horizon of the black hole. This reflection is interesting as it is equivalent to having a particle escape from the black hole \cite{kuchiev2003reflection}, as would be seen with Hawking radiation. In most field theories these reflected, or escaping, particles exhibit a well defined resonance which is unique to the particle and the black hole from which it escaped. We call this resonance the Quasi Normal Modes (QNM) of the particle, and this resonance can uniquely describe the parameters of the black hole from which the particle is emitted.
 
\par This proceedings will focus on the absorption probability of spin-3/2 particles, where our motivation for studying this particle is two fold: Firstly from a theoretical point, due to the scarcity of work focusing on these particles. Secondly, in many super gravity theories gravity is strongly coupled to the massless Rarita-Schwinger fields, spin-3/2 fields, which act as a source of torsion and curvature on the space time \cite{das1976gauge,grisaru1977supergravity}. 

\par Using the method developed by Unruh we will present the equations required to determine the absorption probabilities for the low energy cases \cite{unruh1976absorption}. We will then use the WKB method developed by Iyer and Will, to determine a solution for the more general case \cite{iyer1987black}.

\section{$N$-dimensional Schwarzschild}

\subsection{Potential function}

\par In order to implement the Unruh method we must use a Klein-Gordon equation to describe our particle. We have chosen to use the massless Rarita-Schwinger equation,
\begin{equation}\label{Eq:Rarita}
\gamma^{\mu\nu\alpha}\nabla_{\nu}\Psi_{\alpha}= 0 \; ,
\end{equation}
where $\gamma^{\mu\nu\alpha}= \gamma^{\mu}\gamma^{\nu}\gamma^{\alpha} + \gamma^{\mu}g^{\nu \alpha} - \gamma^{\nu}g^{\mu\alpha} + \gamma^{\alpha}g^{\mu\nu}$. We also need to describe the space time this is being calculated in using the $N$-dimensional metric,
\begin{equation}\label{Eq:Metric}
ds^{2} = -f(r)dt^{2} + f(r)^{-1}dr^{2} + r^{2}d\Omega_{N-2}^{2} \; ,
\end{equation}
where $d\Omega_{N-2}^{2} = d\theta^{2} + \sin\theta d\theta_{N-3}^{2}$ is the metric describing the $N-2$ sphere. Using Eq.(\ref{Eq:Rarita}) and Eq.(\ref{Eq:Metric}) we can derive our equations of motion. 

\par In the case of $N$-dimensional black holes we have two sets of equations of motion, and therefore two potentials. This is because we have both spinors and spinor-vectors in the $N$-dimensional space time, spinor-vectors have both TT-eigenmodes and non-TT eigenmodes. Please refer to Ref.\cite{Ndimensionalpaper} for a complete description. In the interests of brevity we will focus only on the spinor eigenmodes as the method and results are similar for the spinor-vector eigenmodes.

\par In solving for our potential we begin by determining the equations of motion of our particle. We then perform a conformal transformation on these equations of motion to obtain a Schr\"{o}dinger like equation, which we have called a radial equation. From this radial equation we can determine our potential function. The radial equations are given as,
\begin{eqnarray}
-\frac{d^{2}}{dr_{*}^{2}}\tilde{\Phi}_{1}+V_{1}\tilde{\Phi}_{1}
=\omega^{2}\tilde{\Phi}_{1}\ \ \ ,\ \ \ -\frac{d^{2}}{dr_{*}^{2}}\tilde{\Phi}_{2}+V_{2}\tilde{\Phi}_{2}
=\omega^{2}\tilde{\Phi}_{2} \;.\label{SL}
\end{eqnarray}
where $r_{*}$ is the tortoise coordinate, given as $d/dr_{*} = f(r) d/dr$ and $f(r) = 1 - (2M/r)^{N-3}$, and our potential is \cite{Ndimensionalpaper}
\begin{equation}\label{Non-TTPotential}
V_{1,2} = \pm f \frac{dW}{dr} + W^{2} ,
\end{equation}
with
\begin{equation}\label{EqWNTT}
W = \frac{|\bar{\lambda}|\sqrt{f}}{r}\left( \frac{\left(\frac{2}{N-2} \right)^{2}|\bar{\lambda}|^{2} - 1- \frac{N-4}{N-2}\left(\frac{2M}{r} \right)^{N-3}}{\left(\frac{2}{N-2}\right)^{2}|\bar{\lambda}|^{2} - f}\right).
\end{equation}
In solving for the absorption probabilities we will work with $V_{1}$ and set $M=1$.

\subsection{The Unruh method}

\par There are three regions around the black hole that we must consider. Namely the near region, where $f(r) \rightarrow 0$, the central region, where $V(r) \gg \omega$, and the far region, where $f(r) \rightarrow 1$. Approximations are obtained for each of the regions and then coefficients are determined by comparing and evaluating the solutions at the boundaries.

\subsubsection{Near region}

\par In the near region $f(r) \rightarrow 0$, Eq.(\ref{SL}) becomes
\begin{equation}
\left(\frac{d^{2}}{dr_{*}^{2}} + \omega^{2} \right)\widetilde{\Phi}_{I} = 0 \; ,
\end{equation} 
which has a solution
\begin{equation}
\widetilde{\Phi}_{I} = A_{I}e^{i\omega r_{*}}.
\end{equation}

\subsubsection{Central region}

\par In this region we have that $V(r) \gg \omega$ and hence Eq.(\ref{SL}) becomes
\begin{equation}
\left(\frac{d}{dr_{*}} + W \right)\left(\frac{d}{dr_{*}} - W \right)\widetilde{\Phi}_{II} = 0 \; .
\end{equation}
Defining $H$ as 
\begin{equation}\label{EqH}
H = \left(\frac{d}{dr_{*}} - W \right)\widetilde{\Phi}_{II} \; ,
\end{equation}
the solution of Eq.(\ref{EqH}) is
\begin{equation}\label{SolH}
H = B_{II}\left(\frac{1 + \sqrt{f}}{1-\sqrt{f}} \right)^{\frac{j}{N-3}+ \frac{1}{2}}\left(\frac{\left(\frac{2}{N-2} \right)\left(j + \frac{N-3}{2} \right)- \sqrt{f}}{\left(\frac{2}{N-2} \right)\left(j + \frac{N-3}{2} \right) + \sqrt{f}} \right).
\end{equation}
Substituting Eq.(\ref{SolH}) into Eq.(\ref{EqH}), we have a first order differential equation with solution,
\begin{equation} 
\begin{aligned}
\widetilde{\Phi}_{II} = A_{II}\left(\frac{1 + \sqrt{f}}{1-\sqrt{f}}\right)^{\frac{j}{N-3} + \frac{1}{2}} \left(\frac{\left( \frac{2}{N-2}\right)\left(j + \frac{N-3}{2}\right)-\sqrt{f}}{\left(\frac{2}{N-2} \right)\left(j + \frac{N-3}{2}\right)} + \sqrt{f}\right) + B_{II}\Psi \; ,
\end{aligned}
\end{equation}
where $\Psi$ is 
\begin{equation} 
\begin{aligned}
\Psi  = & \left(\frac{1 + \sqrt{f}}{1 -\sqrt{f}} \right)^{\frac{j}{N-3} + \frac{1}{2}}\left(\frac{\left( \frac{2}{N-2}\right)\left(j + \frac{N-3}{2}\right)-\sqrt{f}}{\left(\frac{2}{N-2} \right)\left(j + \frac{N-3}{2}\right)} + \sqrt{f}\right)\\
& \times\left[\int\limits^{r}\frac{1}{f}\left(\frac{1 - \sqrt{f}}{1 +\sqrt{f}} \right)^{\frac{2j}{N-3} + 1}\left(\frac{\left( \frac{2}{N-2}\right)\left(j + \frac{N-3}{2}\right)+\sqrt{f}}{\left(\frac{2}{N-2} \right)\left(j + \frac{N-3}{2}\right)} - \sqrt{f}\right)^{2}dr' \right].
\end{aligned}
\end{equation}

\subsubsection{Far region}

\par In the far region $f(r) \rightarrow 1$, Eq.(\ref{SL}) becomes
\begin{equation}\label{Eq:FarnonTT} 
\begin{aligned}
\frac{d^{2}}{dr_{*}^{2}}\widetilde{\Phi}_{III} - \left[\frac{\left(\left(j + \frac{D-4}{2} \right)^{2} - \frac{1}{4} \right)}{r^{2}} - \omega^{2} \right]\widetilde{\Phi}_{III} = 0 \; .
\end{aligned}
\end{equation}
In this region $r_{*} \sim r$, the solution can be expressed as a Bessel function
\begin{equation} 
\begin{aligned}
\widetilde{\Phi}_{III} = A_{III}\sqrt{r}J_{j+\frac{N-4}{2}}(\omega r) + B_{III}\sqrt{r}N_{j+\frac{N-4}{2}}(\omega r) \; .
\end{aligned}
\end{equation}
Setting the amplitude of $\widetilde{\Phi}_{III}$ to one at spatial infinity, comparing the solutions of region $I$ and $II$, and regions $II$ and $III$, we find that the probability amplitude is
\begin{equation} 
\begin{aligned}
\left| A_{j}(\omega)\right|^{2} = 4\pi C^{2} \omega^{2j +N-3}\left(1 + \pi C^{2} \omega^{2j+N-3} \right)^{-2} \approx 4 \pi C^{2}\omega^{2j+N-3} \; .
\end{aligned}
\end{equation}
The constant
\begin{equation} 
\begin{aligned}
C = \frac{1}{2^{\frac{N-1}{N-3}j +\frac{N-1}{2}}\Gamma(j+\frac{N-2}{2})}\left(\frac{j+ \frac{2N-5}{2}}{j - \frac{1}{2}}\right),
\end{aligned}
\end{equation}
with $\Gamma$ denoting the gamma function and $\omega$ being less than 1. This restriction on $\omega$ means that we require another method in order to determine the absorption probability of our entire energy spectrum of particles. We have chosen to use the WKB method in order to do this.

\subsection{WKB method}

\par It is more convenient in this case to write $x = \omega r$ and take $Q(x) = \omega^{2} - V$. Eq.(\ref{SL}) then becomes
\begin{equation} 
\begin{aligned}
\left(\frac{d^{2}}{dr_{*}^{2}} + Q \right)\widetilde{\Phi}_{1} = 0 \; .
\end{aligned}
\end{equation}
For energies $\omega \ll V$ it is sufficient to use the first order WKB approximation \cite{cho2005wkb},
\begin{equation}\label{FirstOrderWKB}
\begin{aligned}
|A_{j}| = exp\left[-2 \int\limits_{x_{1}}^{x_{2}}\frac{dx'}{f(x')}\sqrt{-Q(x')} \right],
\end{aligned}
\end{equation}
where $x_{1}$ and $x_{2}$ are turning points, with $Q(x_{1},x_{2}) = 0$ or $V_{x_{1},x_{2}}=\omega^{2}$, for a given energy $\omega$ and potential $V$. When $\omega^{2} \sim V$ the formula in Eq.(\ref{FirstOrderWKB}) is no longer convergent. For this energy we will need to use a higher order WKB approximation. We use the third order method developed by Iyer and Will \cite{iyer1987black}. The absorption probability is given as 
\begin{equation} \label{Absorption}
\begin{aligned}
|A_{j}(\omega)|^{2} = \frac{1}{1+ e^{2S(\omega)}} \; ,
\end{aligned}
\end{equation}
where 
\begin{equation} \label{SW}
\begin{aligned}
S(\omega) = &  - \pi k^{-1/2}\left[\frac{1365}{2048}b_{3}^{4} - \frac{525}{256}b_{3}^{2}b_{4} + \frac{85}{128}b_{4}^{2} + \frac{85}{128}b_{4}^{2} + \frac{95}{64}b_{3}b_{5} - \frac{25}{32}b_{6} \right]z_{0}^{2}\\
& + \pi k^{-1/2}\left[\frac{3}{16}b_{4} -\frac{7}{64}b_{3}^{2} \right]+ \pi k^{1/2}\left[\frac{1}{2}z_{0}^{2} + \left(\frac{15}{64}b^{2}_{3} - \frac{3}{16}b_{4} \right)z_{0}^{4} \right]  \\
& + \pi k ^{1/2}\left[\frac{1155}{2048}b_{3}^{4} - \frac{315}{256}b_{3}^{2}b_{4} + \frac{35}{128}b^{2}_{4} + \frac{35}{64}b_{3}b_{5} - \frac{5}{32}b_{6} \right]z_{0}^{6} \; , \\
\end{aligned}
\end{equation}
for $z_{0}^{2},b_{n}$ and $k$ defined by the components of the Taylor series expansion of $Q(x)$ near $x_{0}$. That is,
\begin{equation} 
\begin{aligned}
Q & = Q_{0} + \frac{1}{2}Q_{0}''z^{2} + \sum\limits_{n=3} \frac{1}{n!}\left( \frac{d^{n}Q}{dx^{n}}\right)_{0}z^{n} = k\left[z^{2} - z^{2}_{0} + \sum\limits_{n=3}b_{n}z^{n} \right],
\end{aligned}
\end{equation}
\linebreak
and
\begin{equation} 
\begin{aligned}
z & = x - x_{0}; \: z_{0}^{2} \equiv -2\frac{Q_{0}}{Q_{0}''} ;\:k  \equiv \frac{1}{2}Q_{0}'' ; \:  b_{n}  \equiv \left(\frac{2}{n!Q_{0}''} \right)\left(\frac{d^{n}Q}{dx^{n}} \right)_{0}.\\
\end{aligned}
\end{equation}
where subscript $0$ denotes the maximum and the primes denote derivatives.

\section{Results and conclusions}

\begin{figure}[ht]
\begin{subfigure}{0.5\textwidth}
\includegraphics[scale=0.45]{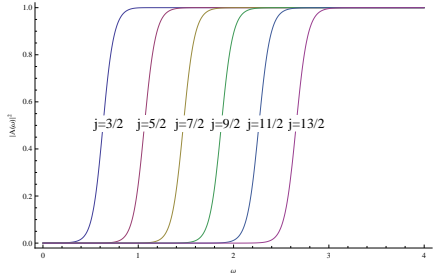}
\caption{Absorption probability and particle energy \\ 
for 4-dimensional Schwarzschild black hole.}
\label{non-TT 4D}
\end{subfigure}
\begin{subfigure}{0.5\textwidth}
\includegraphics[scale=0.45]{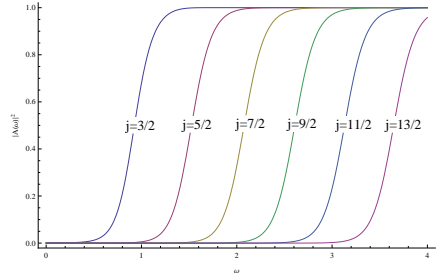}
\caption{Absorption probability and particle energy \\
for 5-dimensional Schwarzschild black hole.}
\label{non-TT 5D}
\end{subfigure}
\begin{subfigure}{0.5\textwidth}
\includegraphics[scale=0.45]{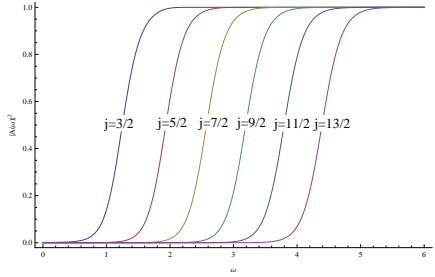}
\caption{Absorption probability and particle energy \\
for 6-dimensional Schwarzschild black hole.}
\label{non-TT 6D}
\end{subfigure}
\begin{subfigure}{0.5\textwidth}
\includegraphics[scale=0.45]{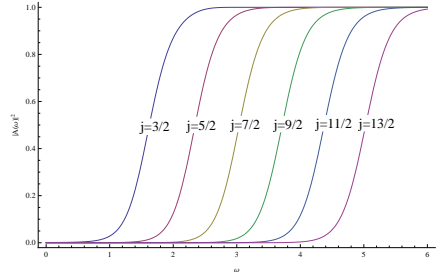}
\caption{Absorption probability and particle energy \\
for 7-dimensional Schwarzschild black hole.}
\label{non-TT 7D}
\end{subfigure}
\caption{Absorption probabilities associated to particles of spin-3/2 with $j$ values ranging from 3/2 to 13/2 near Schwarzschild black holes of dimension 4 to 7.}
\end{figure}
\begin{figure}[ht]
\begin{subfigure}{0.5\textwidth}
\includegraphics[scale=0.45]{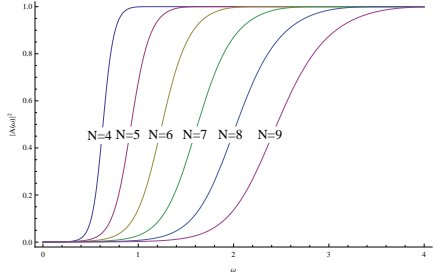}
\caption{Absorption probability and particle \\
 energy for j=3/2 on various \\
 dimensional Schwarzschild black holes.}
\label{fig: j=3/2}
\end{subfigure}
\begin{subfigure}{0.5\textwidth}
\includegraphics[scale=0.45]{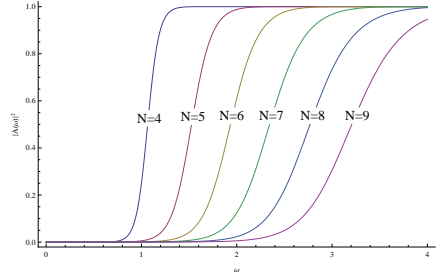}
\caption{Absorption probability and particle \\
 energy for j=5/2 on various \\
 dimensional Schwarzschild black holes.}
\label{fig: j=5/2}
\end{subfigure}
\begin{subfigure}{0.5\textwidth}
\includegraphics[scale=0.45]{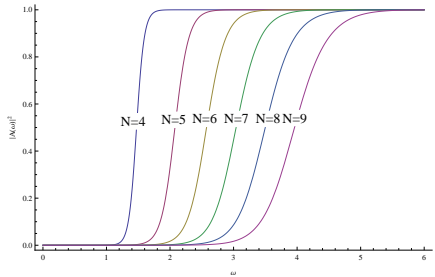}
\caption{Absorption probability and particle \\
 energy for j=7/2 on various \\
 dimensional Schwarzschild black holes.}
\label{fig: j=7/2}
\end{subfigure}
\begin{subfigure}{0.5\textwidth}
\includegraphics[scale=0.45]{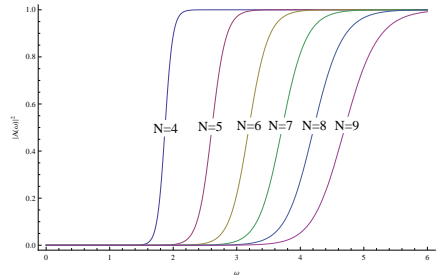}
\caption{Absorption probability and particle \\
 energy for j=9/2 on various \\
 dimensional Schwarzschild black holes.}
\label{fig: j=9/2}
\end{subfigure}
\caption{Absorption probability for particles of $j=3/2$ to $j=9/2$ for various dimensional Schwarzschild black holes.}
\label{fig:3/2to13/2}
\end{figure}
\par The results for our absorption probabilities, for spinors, are given in Figs. 1 and 2. Although we have not explicitly given the effective potential for the spinor-vectors, they are derived in the same way as the potential for our spinors. It is interesting to note that the potential for our spinor-vectors can be reduced to the potential for spin-1/2 particles \cite{Ndimensionalpaper,cho2007split}. Therefore if we look at the absorption probability for spin-1/2 particles and our spin-3/2 spinor vector particles we find that their absorption probabilities are consistent with each other.
 
\par Setting $N=4$ in our potential yields the same potential as we have obtained in Ref. \cite{chen2015gravitino}. Looking at Figs. \ref{non-TT 4D} to \ref{non-TT 7D} we see that increasing the excitation level of our particle increases the minimum energy required for absorption to take place. This means that more excited particles will be emitted from the black hole with a higher energy. Next, looking at Figs. \ref{fig: j=3/2} and \ref{fig: j=5/2} we see that higher dimensional black holes will emit more energetic particles, since the minimum energy required for absorption increases with an increasing number of dimensions. 


\section*{Acknowledgements}

\noindent We would like thank Wade Naylor for his useful discussions during the production of this work. ASC and GEH are supported in part by the National Research Foundation of South Africa (Grant No: 91549). CHC and HTC are supported in part by the Ministry of Science and Technology, Taiwan, ROC under the Grant No. NSC102-2112-M-032-002-MY3, and by the National Center for Theoretical Sciences (NCTS). GEH would like to thank CHC and HTC for their hospitality during his visit to Tamkang university, where part of this work was completed. 

\section*{References}

\bibliography{iopart-num}

\end{document}